\documentclass[runningheads]{llncs}
\usepackage[T1]{fontenc}
\usepackage{graphicx,verbatim}
\usepackage{bm}
\usepackage{shortcutsg}
\usepackage{graphicx}
\usepackage{subfigure}
\usepackage{booktabs} % for professional tables
\usepackage{rotating}
\usepackage{url}
\newcommand{\modelname}[1]{Morph}
\newcommand{\modelnameldm}[1]{\modelname{}LDM}
\newcommand{\modelnameldmcond}{\modelnameldm{}\textsuperscript{c}}
\begin{document}
\title{Generating Novel Brain Morphology by Deforming Learned Templates}
%
% Removed for anonymized MICCAI 2025 submission
\author{Alan Q. Wang\inst{1} 
\and
Fangrui Huang\inst{1}
\and
Bailey Trang\inst{1}
\and
Wei Peng\inst{1}
\and
Mohammad Abbasi\inst{1}
\and
Kilian M. Pohl\inst{1}
\and
Mert R. Sabuncu\inst{2}
\and
Ehsan Adeli\inst{1}
}
\authorrunning{A. Wang et al.}
% First names are abbreviated in the running head.
% If there are more than two authors, 'et al.' is used.
%
\institute{Stanford University, Stanford, CA 94305, USA
\and
Cornell Tech and Weill Cornell Medicine, New York, NY 10044, USA
\email{alanqw@stanford.edu}
}
% \author{Alan Wang, Fangrui Huang, Bailey Trang, Wei Peng, Mohammad Abbasi, Kilian Pohl, Mert Sabuncu, Ehsan Adeli}  %% Added for anonymized MICCAI 2025 submission
% \authorrunning{Wang et al.}
% \institute{Anonymized Affiliations \\
%     \email{email@anonymized.com}}

\maketitle              % typeset the header of the contribution
\begin{abstract}
Designing generative models for 3D structural brain MRI that synthesize morphologically-plausible and attribute-specific (e.g., age, sex, disease state) samples is an active area of research. %offering the potential for augmenting limited training data, generating counterfactuals, and increasing interpretability. 
Existing approaches based on frameworks like GANs or diffusion models synthesize the image directly, which may limit their ability to capture intricate morphological details. 
% Some works have explored leveraging GAN and diffusion-based approaches for synthesizing deformation maps, which restricts the generator to focus on intricate morphological details rather than the entire image. 
% These deformation maps are then applied to a template image.
% However, these approaches typically choose a fixed template or precompute the deformation maps for training the generator.
In this work, we propose a 3D brain MRI generation method based on state-of-the-art latent diffusion models (LDMs), called \modelnameldm{}, that generates novel images by applying synthesized deformation fields to a learned template.  % we propose a method, called \modelname{}, for generating morphology by synthesizing spatial transformations, which are then applied to a learned, universal template to produce a novel image. 
% The fixed template is not synthesized but is the output of a template generator which is optimized concurrently with the morphology generator.
% Applied to state-of-the-art latent diffusion models (LDMs), our approach turns the latent representation of an MRI and a learnable template into a dense deformation field. % through the latent diffusion and iterative denoising process.  
%Our approach differs from LDMs in that the latent embeddings that the diffusion model is trained on is not learned by a reconstruction-based autoencoder.
Instead of using a reconstruction-based autoencoder (as in a typical LDM), our encoder outputs a latent embedding derived from both an image and a learned template that is itself the output of a template decoder;
% The encoder takes in both an image and the learned template and outputs a latent embedding.
this latent is passed to a deformation field decoder, whose output is applied to the learned template. 
A registration loss is minimized between the original image and the deformed template with respect to the encoder and both decoders.
% with respect to the autoencoder and template generator network. 
Empirically, our approach outperforms generative baselines on metrics spanning image diversity, adherence with respect to input conditions, and voxel-based morphometry.

Our code is available at \url{https://github.com/alanqrwang/morphldm}.

% Designing generative models for 3D structural brain MRI that synthesize morphologically-plausible and attribute-specific (e.g., age, sex, disease state) samples is an active area of research. Existing approaches based on frameworks like GANs or diffusion models synthesize the image directly, which may limit their ability to capture intricate morphological details. In this work, we propose a 3D brain MRI generation method based on state-of-the-art latent diffusion models (LDMs), called \modelnameldm{}, that generates novel images by synthesizing deformation fields applied to a learned template. Instead of using a reconstruction-based autoencoder (as in LDM), our encoder outputs a latent embedding derived from both an image and a \textit{learned template} that is itself the output of a template decoder; this latent is passed to a deformation field decoder, whose output is applied to the learned template. A registration loss is minimized between the original image and the deformed template with respect to the encoder and both decoders. Empirically, our approach outperforms generative baselines on metrics spanning image diversity, adherence with respect to input conditions, and voxel-based morphometry. Our code will be made available upon acceptance.

% Authors must provide keywords and are not allowed to remove this Keyword section.

\end{abstract}

\begin{figure*}[t]
  \centering
   \includegraphics[width=0.95\linewidth]{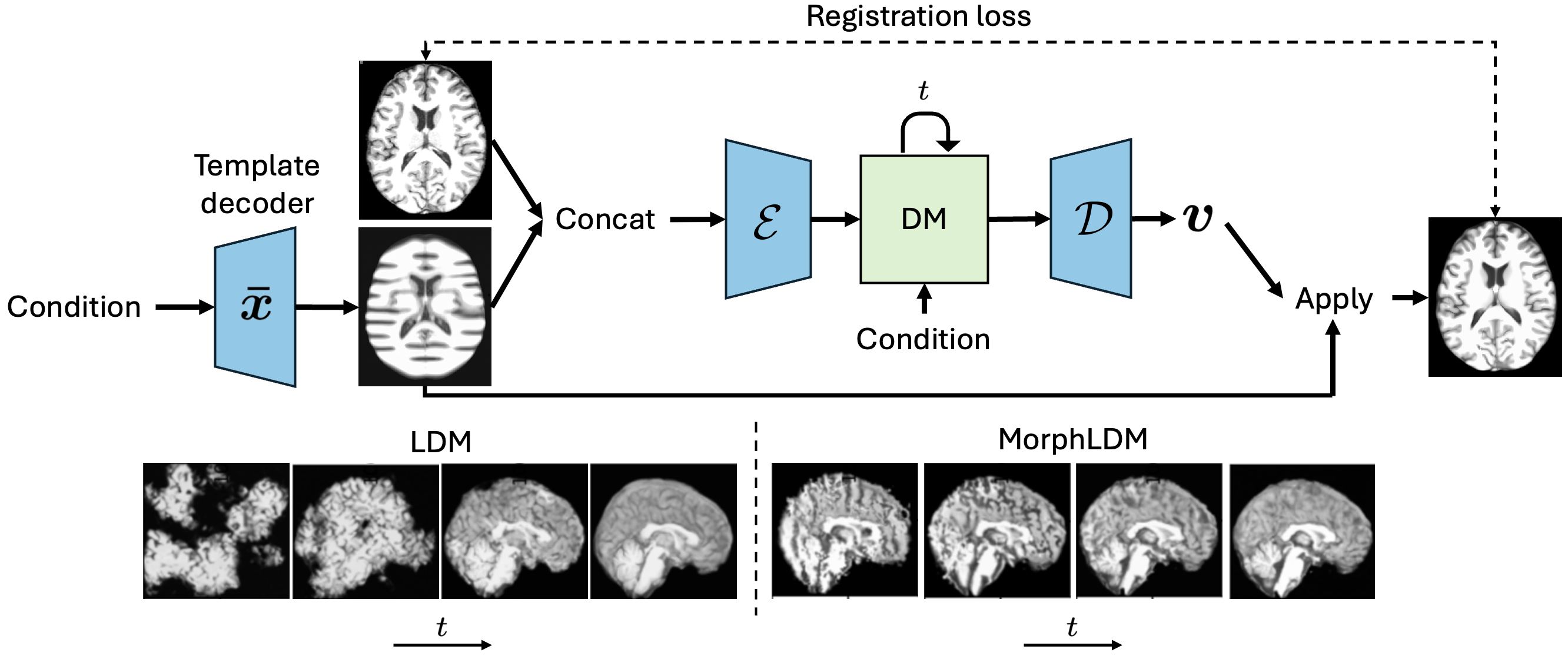}

   \caption{(top) Graphical depiction of \modelnameldm{}.
   The encoder takes in an image and a template that is the output of a template decoder, concatenated along the channel dimension.
   The decoder outputs a deformation field $\bm{v}$ that is applied to the template.
   Encoder-decoder networks in blue are trained in the first stage on a registration loss (similar to LDM). The diffusion model (DM) in green is trained on the learned latents in the second stage.  (bottom) Visualization of denoising steps $t \in [T]$ of LDM (left) and \modelnameldm{} (right). While LDM's decoder maps from a synthesized latent to an image directly, \modelnameldm{}'s decoder maps to a synthesized deformation field, which is subsequently applied to the learned template.
   % Hence, even in the noisiest steps, the generated images maintain the morphological characteristics of a brain.
   }
   \label{fig:arch}
\end{figure*}

\section{Introduction}
Morphology is an important characteristic of neuroimaging, known to be related to attributes and/or phenotypes such as age, sex, and disease state~\cite{madan2017advances,wang2021independent}. 
A growing literature aims to design generative models for structural brain magnetic resonance imaging (MRI), which synthesize realistic morphology and are specific to conditioned attributes such as age and sex~\cite{PENG2024103325,pinaya2022brainimaginggenerationlatent}. 
These models offer the potential to augment limited training data~\cite{pombo2023equitablemodelling}, generate counterfactuals~\cite{jeanneret2022diffusionmodelscounterfactualexplanations}, and increase interpretability~\cite{wang2024frameworkinterpretability}.

In computer vision, generative models have found success in natural image domains, driven largely by GANs~\cite{fernandez2024multipathological,goodfellow2014generativeadversarialnetworks,yi2019ganinmedicalimaging} and diffusion models~\cite{ho2020denoisingdiffusionprobabilisticmodels,rombach2022highresolutionimagesynthesislatent}.
These models work by learning the data distribution $p(\bm{x})$ from which the data is assumed to be sampled.
Although highly general and capable of capturing the variety inherent in naturalistic images, this approach applied to neuroimaging may fail to capture intricacies and subtleties crucial to morphology, especially in the context of high-dimensional 3D volumetric data and data-limited scenarios.
% Notably, these methods generate all features of the image (intensity, structure, etc.) in one go.

% One way to capture morphology is via spatial deformations encoded as dense deformation fields on the 3D grid. 
Specific to neuroimaging, previous work has shown that much of the variability between individual anatomies can be captured by geometric and topological changes~\cite{christensen1997volumetric,dalca2019learningconditionaldeformable}.
This motivates a line of work that, instead of directly synthesizing images, synthesizes dense deformation fields that are applied to a template image.
% Leveraging neural networks to learn these deformation fields is common in neuroimaging tasks like image registration~\cite{Balakrishnan_2019,wang2023robustinterpretabledeeplearning} and template construction~\cite{dalca2019learningconditionaldeformable,dey2022generativeadversarialregistrationimproved,starck2024diffdefdiffusiongenerateddeformationfields}.
However, in addition to using less powerful generative models like GANs, prior work either precomputes deformation fields using an expensive registration method~\cite{chong2021synthesis,zhuo2024diffuseregdenoisingdiffusionmodel} or fixes the template, thus biasing and limiting the model's ability to adequately capture the image distribution~\cite{chong2021synthesis,kim2022diffusiondeformablemodel4d,zheng2024deformationrecoverydiffusionmodeldrdm}.
% Our work is perhaps most similar to Wang et al.~\cite{wang2023gendeflearninggenerativedeformation}, who address naturalistic video generation by learning both a canonical space and frame-level deformation field applied to the canonical space. 

In this work, we propose a 3D brain MRI generation method based on state-of-the-art latent diffusion models (LDMs), called \modelnameldm{}, that generates novel images by applying synthesized deformation fields to a \textit{learned} template.
% That is, the template is the output of a separate decoder that is learned. 
% During a first stage of training, the encoder of our LDM takes as input an image concatenated with a template that is outputted by a template decoder, and outputs a latent embedding that is passed to a deformation field decoder. 
\modelnameldm{} differs from LDMs in the design of the encoder/decoder.
First, a learned template is outputted by a template decoder, optionally conditioned on image-level attributes.
Then, an encoder takes in both an image and the template and outputs a latent embedding;
this latent is passed to a deformation field decoder, whose output deformation field is applied to the template. 
Finally, a registration loss is minimized between the original image and the deformed template with respect to the encoder and both decoders.
% This deformation field is applied to the template, and a registration loss is optimized between the input image and the deformed template.
Subsequently, a diffusion model is trained on these learned latent embeddings. %, similar to standard LDM training.
% The template itself can be conditioned on attributes by passing these attributes as input to the template decoder.
% Specifically, instead of producing a reconstruction, our autoencoder takes as input an image concatenated with a learnable template and outputs a dense deformation field. 
% This deformation field is applied to the learnable template, and a registration loss is minimized between the input image and deformed image.

To synthesize an image, \modelnameldm{} generates a novel latent in the same way as standard LDMs. 
The decoder maps this latent to its corresponding deformation field, which is subsequently applied to the learned template.
We find that synthetic samples generated by our approach not only outperform powerful baselines on standard image generation metrics like FID, but also better capture input conditions known to be related to morphology, including age and sex.
Additionally, we perform a voxel-based morphometry~\cite{Whitwell9661} analysis and find that our models generate more realistic structures in terms of regional volume compared to baselines.

To the best of our knowledge, \modelnameldm{} is the first MRI diffusion model that synthesizes deformation fields applied to a learned template. % work to propose an MRI generation method based on state-of-the-art LDMs that .
Our approach is efficient and data-driven in the sense that both the deformation fields and the template are learned simultaneously, without requiring precomputation of deformation fields or making a specific choice for the template.
% \subsubsection{Relation to Prior Work}
% Prior work has explored generating deformation fields as a means of synthesizing novel images.

\section{Background}
\subsection{Brain MRI Generation}
\label{sec:ldm}
% Designing generative models for brain magnetic resonance imaging (MRI) is an active area of research.
Generative adversarial networks (GANs) and their extensions have been widely explored in the context of brain MRIs, including using Wasserstein GANs~\cite{gulrajani2017improvedtrainingwassersteingans}, incorporating variational autoencoders (VAEs) and code discriminators~\cite{kwon2019braingan}, and leveraging segmentations~\cite{fernandez2024multipathological}.
Autoregressive (AR) approaches generate samples in an autoregressive fashion and require discrete tokenization of images, typically done with a vector-quantized (VQ)-VAE~\cite{esser2021tamingtransformershighresolutionimage,tudosiu2022autoregressive}.
% which synthesize realistic morphology and which are specific to conditioned attributes like age and sex is a subject of much research ~\cite{pinaya2022brainimaginggenerationlatent,PENG2024103325}, and offers the potential for augmenting limited training data, increasing the equitability of training data~\cite{pombo2023equitablemodelling}, generating counterfactuals~\cite{jeanneret2022diffusionmodelscounterfactualexplanations}, and increasing interpretability~\cite{wang2024frameworkinterpretability}.

More recently, latent diffusion models (LDMs) have become the state-of-the-art generative model for brain MRIs~\cite{PENG2024103325,pinaya2022brainimaginggenerationlatent}, and are trained in two steps.
In the first stage, an autoencoder $\Dcal(\Ecal(\bm{x})) = \Dcal(\bm{z})$ is trained on a reconstruction loss $\Lcal_{sim}$ to learn a latent space. 
% such that the encoder $\Ecal$ learns to map $\bm{x}$ to a latent embedding $\bm{z}$.
% Typically, this autoencoder is trained to optimize a reconstruction loss.
% Let $\mathcal{E}_\theta$ and $\mathcal{D}_\phi$ denote the image encoder and decoder parameterized by $\theta$ and $\phi$, respectively.
% The reconstruction loss is:
% \begin{equation}
    % \argmin_{\theta, \phi} \EE_{\bm{x}} \ \Lcal(\Dcal_\phi(\Ecal_\theta(\bm{x})), \bm{x}).
    % \label{eq:recon_loss}
% \end{equation}
% $\Lcal$ denotes a reconstruction loss, where common choices are mean-squared-error (MSE), mean absolute error (MAE), and/or perceptual losses~\cite{johnson2016perceptuallossesrealtimestyle,czolbe2021lossfunctiongenerativeneural}.
% Adversarial losses are common as well~\cite{pinaya2022brainimaginggenerationlatent,PENG2024103325}.
In the second stage, a diffusion model is trained on the latent space defined by $\bm{z} = \Ecal(\bm{x})$; specifically, a network $\epsilon_\omega$ is trained to perform time-conditioned denoising of the latent embeddings:
\begin{equation}
\argmin_\omega \EE_{\Ecal(\bm{x}),\epsilon \sim \Ncal(0,1),t \in [T]} ||\epsilon - \epsilon_\omega(\bm{z}_t,t, \bm{c})||^2_2,
\label{eq:ldm}
\end{equation}
% Omitting $\bm{c}$ corresponds to an unconditional diffusion model.
where $\EE$ is expected value, $\Ncal(0, I)$ is the standard normal distribution, and $\bm{c}$ is any conditioning information associated with the image.\footnote{Typically, $\epsilon_\omega$ is implemented as a time-conditional UNet~\cite{rombach2022highresolutionimagesynthesislatent}, where the conditioning information $\bm{c}$ is included via cross-attention with the intermediate feature maps.}

\subsection{Templates}
\label{sec:deformable_templates}
Deformable templates, or atlases, are common in computational anatomy. 
Research is dedicated to constructing unbiased templates such that the deformation fields from these templates to individual images can be analyzed to understand population variability~\cite{BLAIOTTA2018117,christensen1997volumetric,GANSER20043}. 
% Classically, the template is usually constructed through an iterative but expensive procedure based on a collection of images or volumes~\cite{thompson2000templates}. 
% The images can be subdivided to build multiple atlases conditioned on a specific subpopulation. 
% For example, in neuroimaging, some methods
% build different templates for different age groups, requiring rigid discretization of the population and prohibiting each template from using all information across the collection. 
% Images can also be clustered and a template optimized for each cluster, requiring a pre-set number of clusters [63].
% Specialized methods have also been developed that tackle a particular variability of interest. For
% example, spatiotemporal brain templates have been developed using specialized registration pipelines
% and explicit modelling of brain degeneration with time [22, 31, 48], requiring significant domain
% knowledge, manual anatomical segmentations, and significant computational resources. 
More recently, template learning approaches use neural networks to learn templates that are data-driven and attribute-specific~\cite{dalca2019learningconditionaldeformable,dey2022generativeadversarialregistrationimproved,starck2024diffdefdiffusiongenerateddeformationfields}. 
These approaches aim to build decoders that learn templates conditioned on specified attributes.
% In computer vision more broadly, templates have been used as intermediate feature layers for improving object recognition~\cite{bonde2015templatenetdepthbasedobjectinstance} and 3D surface and shape generation~\cite{groueix2018atlasnetpapiermacheapproachlearning,kurenkov2017deformnetfreeformdeformationnetwork}.
% Specifically, our strategy learns a single network that levarges shared information across the entire dataset and can output different templates as a
% function of sets of attributes, such as age, sex, and disease state. 
% The conditional function learned by
% these models generate unbiased population templates for a specific configuration of the attributes.
% Our model can be used to study the population variation with respect to certain attributes it was trained
% on, such as age in neuroimaging. In recent literature on deep probabilistic models, several papers find
% and explore latent axes of important variability in the dataset [4, 15, 30, 33, 43, 51]. Our model can
% also be used to build conditional geometric templates based on such latent information, as we show
% in our experiments. In this case, our model can be seen as learning meaningful image representations
% up to a geometric deformation. However, in this paper we focus on observed (measured) attributes,
% with the goal of explicitly capturing variability that is often a source of confounding.
% Follow-up works have added more sophisticated adversarial losses to improve the quality of learned templates~\cite{dey2022generativeadversarialregistrationimproved}.
In this work, we take inspiration from template learning approaches, where our goal is to learn a universal template concurrently with a deformation field synthesizer to generate novel brain images.

\section{Methods}
\label{sec:formatting}
Let $\bm{x} \in \RR^{C \times L \times W \times H}$ denote a 3D volume sampled from a distribution $p(\bm{x})$.
For MRIs, $\bm{x}$ is a single-channel intensity image where $C=1$.
Let $\bm{z} \sim p(\bm{z})$ denote a sample from some prior distribution (usually a standard normal) and let $\bm{c} \in \Ccal$ denote (optional) conditioning information.
Typical generative approaches aim to model $p(\bm{x})$ directly via a generator $\Gcal$ that takes in a latent sample along with the condition: $\hat{\bm{x}} = \Gcal(\bm{z}, \bm{c})$.

We assume the existence of a template $\bm{\bar{x}} \in \RR^{C \times L \times W \times H}$ such that any $\bm{x} \sim p(\bm{x})$ is related to $\bm{\bar{x}}$ via applying a deformation field $\bm{v} \in \RR^{3 \times L \times W \times H}$: $\bm{x} = \bm{v} \circ \bm{\bar{x}}$. %, where $\circ$ denotes applying the tra spatial transformation.
% $\bm{v}$ is commonly represented as a dense deformation field on the 3D grid.
Under the template assumption, we can recast modeling $p(\bm{x})$ as modeling $p(\bm{v})$, the distribution of deformation fields induced by $\bm{\bar{x}}$.
Then, the output of the generator is a novel deformation field $\hat{\bm{v}} = \Gcal(\bm{z}, \bar{\bm{x}}, \bm{c})$, and the novel image derived from this deformation field is $\hat{\bm{x}} = \hat{\bm{v}} \circ \bar{\bm{x}}$.\footnote{The template assumption is common and well-motivated in the literature~\cite{chong2021synthesis,christensen1997volumetric,dalca2019learningconditionaldeformable}, but notably may be violated when $\bm{x}$ exhibits lesions or large structural abnormalities.}
To prevent any bias due to a specific choice of $\bar{\bm{x}}$, we learn $\bar{\bm{x}}$ concurrently with $\Gcal$ by optimizing a separate template decoder $\bar{\bm{x}}: \Ccal \rightarrow \RR^{C \times L \times W \times H}$, which outputs a deterministic, universal template on which all synthesized deformation fields are applied.
% Specifically, the template decoder maps from an input vector to the template.
In the simplest case, $\bar{\bm{x}}$ can be made unconditional by setting the input vector to be learnable.
We also experiment with adding $\bm{c}$ as an input to the template decoder to learn conditional templates. 
This enables the learned templates to capture intensity or attribute-specific structural detail that can improve overall generation~\cite{dalca2019learningconditionaldeformable,dey2022generativeadversarialregistrationimproved}. 

\subsection{\modelnameldm{}}
% Although, \modelname{} is not restricted to a certain class of generative models, in this work we apply \modelname{} to state-of-the-art LDMs.
In this work, we apply our approach to state-of-the-art LDMs, where the second stage diffusion training remains unchanged (Eq. \eqref{eq:ldm}) and the main changes are in the design and training of the autoencoder.
Fig.~\ref{fig:arch} gives a graphical depiction.
Let $\mathcal{E}_\theta$ and $\mathcal{D}_\phi$ denote the image encoder and decoder parameterized by $\theta$ and $\phi$, respectively.
The input to the autoencoder is $\bm{x}$ and $\bar{\bm{x}}$ concatenated along the channel dimension. 
The encoder output $\bm{z}$ is passed to the decoder, which outputs a deformation field $\bm{v}$ that is applied to the template $\bar{\bm{x}}$ via a differentiable grid-sampler. %~\cite{jaderberg2016spatialtransformernetworks}.
Thus, the objective is
% \begin{equation}
%     \argmin_{\bm{\bar{x}}, \theta, \phi} \EE_{\bm{x}} \left[ \Lcal_{rec}(\Dcal_\phi(\Ecal_\theta(\bm{x}, \bar{\bm{x}})) \circ \bm{\bar{x}}, \bm{x}) + \Rcal(\bm{v}) \right]
%     \label{eq:uncond}
% \end{equation}
% and the conditional template objective is
\begin{equation}
    \argmin_{\bar{\bm{x}}, \theta, \phi} \EE_{\bm{x}} \left[ \Lcal_{sim}(\Dcal_\phi(\Ecal_\theta(\bm{x}, \bar{\bm{x}})) \circ \bar{\bm{x}}, \bm{x}) + \Rcal(\bm{v}) \right],
    \label{eq:cond}
\end{equation}
where $\Rcal(\bm{v})$ denotes a regularization term on the predicted deformation field.
% This term is commonly used in learning-based registration approaches and represents a prior on the deformation fields, thereby restricting the space of permissible registrations~\cite{Balakrishnan_2019}.
In this work, we penalize the magnitude and spatial gradient of the displacement field.
Letting $\bm{u}_{\bm{v}}$ denote the spatial displacement for $\bm{v} = Id + \bm{u}_{\bm{v}}$, where $Id$ is the identity transformation:
$
    \Rcal(\bm{v}) = \alpha ||\bm{u}_{\bm{v}}||_2 + \beta ||\nabla \bm{u}_{\bm{v}}||_2.
$
Following prior work~\cite{pinaya2022brainimaginggenerationlatent,rombach2022highresolutionimagesynthesislatent}, we also impose a slight KL penalty towards a standard normal on the latent space to prevent arbitrarily high-variance latent spaces.\footnote{Many works in this space utilize diffeomorphic constraints~\cite{BLAIOTTA2018117,dalca2019learningconditionaldeformable,pombo2023equitablemodelling}. We found experimentally that this constraint was too strong and led to artifacts in the synthesized samples.}. %, although other regularizers which enforce discrete latent spaces can also be used~\cite{esser2021tamingtransformershighresolutionimage,rombach2022highresolutionimagesynthesislatent}
% Note that $\alpha$ and the weight coefficient for the KL penalty are hyperparameters that control the degree of regularization; these values are tuned via grid search on validation data.

% To do so, instead of optimizing the template image directly on the image grid, we learn a neural network $g_\psi$ which takes as input image metadata and outputs the template representation.
% Note that $g_\psi$ can be made unconditional, for example by replacing $\bm{c}$ with a vector of learnable parameters.
% We experiment with this unconditional template variant to facilitate better comparison with baselines which don't have conditioning information injected in the autoencoder part.

% Intuitively, we argue that this type of model will better model structural and anatomical features, since the generation task is restricted to learning spatial transformations only.

\begin{figure*}[t]
  \centering
   \includegraphics[width=0.92\linewidth]{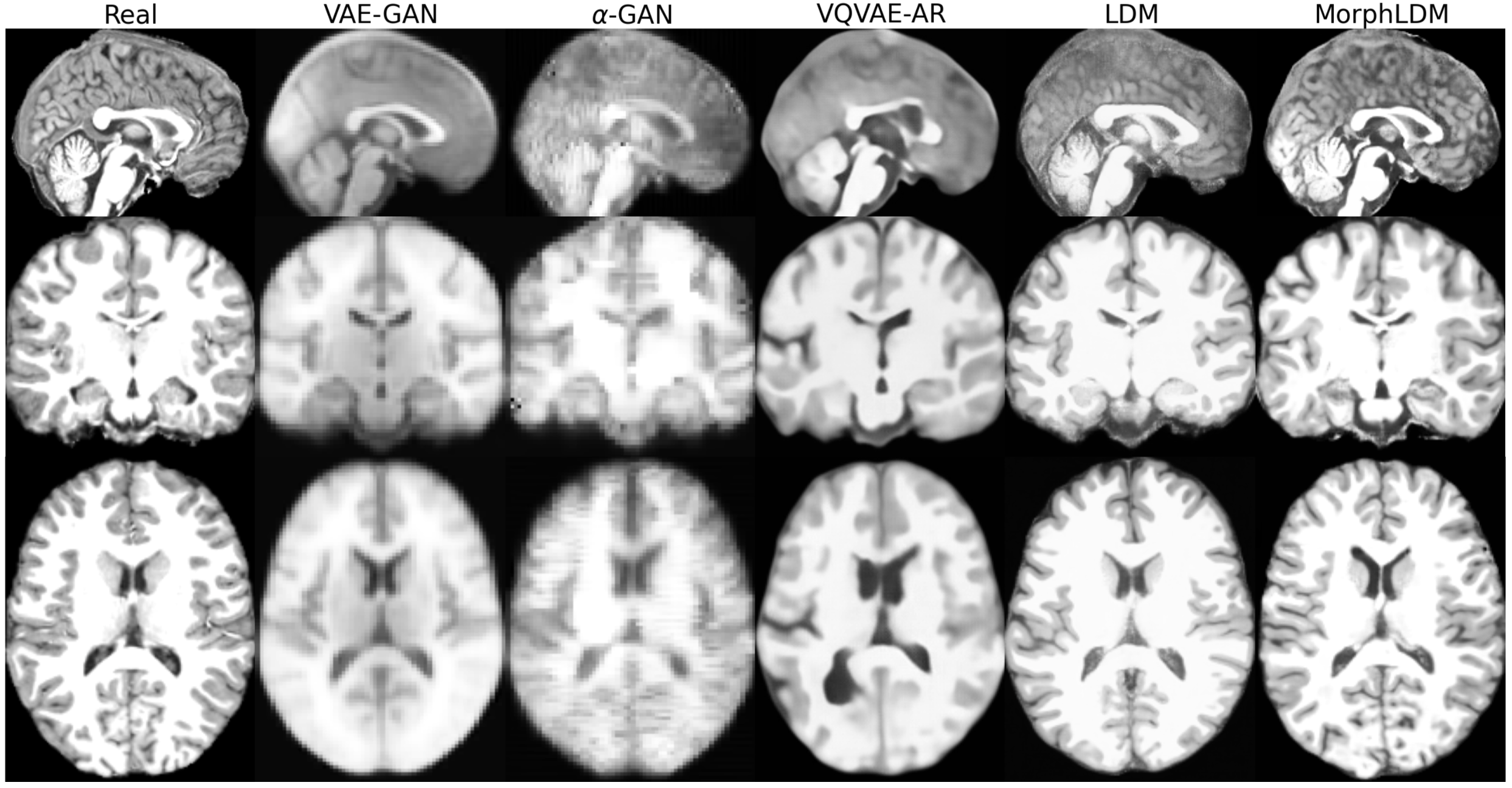}

   \caption{Representative real and synthetic samples for all models shown in the three planes. Age=10, sex=``female''. We observe that \modelnameldm{} has a higher degree of morphological detail compared to baselines, especially in folding patterns in cortical regions.}
   \label{fig:samples}
\end{figure*}

\section{Experiments}
\subsubsection{Dataset.}
We train all models on a large dataset of publicly-available T1-weighted brain MRIs.\footnote{Data are from the Alzheimer's Disease Neuroimaging Initiative (ADNI) \cite{petersen2010alzheimer}, Adolescent Brain Cognitive Development (ABCD) Study~\cite{Karcher2021}, Human Connectome Project (HCP) \cite{VANESSEN201362}, and Parkinson's Progression Markers Initiative (PPMI) \cite{Pulliam2011}.}
For all datasets, we restrict to healthy controls when applicable.
In total, we have 27,066 3D volumes, of which we reserve 21,051 for training and the rest for validation.
Each volume is of resolution 160x192x176 and skullstripped and registered to MNI space.
Age and sex metadata is available for all images. 
Since our dataset is skewed toward younger ages, we uniformly sample ages binned across decades during training.

For generative model evaluation, we generate 1000 samples with conditions linearly spaced between the ages of 5 to 100 and sexes are evenly split between male and female.
For real samples, we collect 1000 validation samples by collecting 500 female validation samples closest in age to the synthetic samples (without replacement), and similarly for male samples. 

\subsubsection{Architectural and Training Details.}
For all LDM-based models, the encoder $\Ecal$ is composed of convolutional blocks with 3 levels of downsampling;
the number of latent channels is 8.
For LDM, $\Dcal$ outputs a 1-channel reconstruction.
For \modelnameldm{}, it outputs a 3-channel output for a 3D deformation field.
The diffusion UNet $\epsilon_\omega$ has intermediate channel sizes of [384, 512, 512], and cross-attention is applied with $\bm{c}$ at the latter two levels.
The architecture of the template decoder $\bar{\bm{x}}$ is similar to the decoder for the deformation field, except that it takes as input a vector with a size equal to the number of conditions $\bm{c}$ and outputs a template $\bar{\bm{x}} \in \RR^{C \times L \times W \times H}$.
% For the grid-sampler, we use trilinear interpolation.

% For $h$, we use 2 convolutional layers with intermediate instance normalization layers and ReLU activations, plus a final convolutional layer with kernel size of 1 which projects down to the number of image channels.
% We found that additional convolutional layers were needed to remove any interpolation artifacts from the grid-sampling procedure.

For training, we use the same reconstruction/similarity loss $\Lcal_{sim}$ for all LDM-based models, which is composed of an L1 loss and an adversarial loss using a patch-based discriminator~\cite{isola2018imagetoimagetranslationconditionaladversarial} with a weight of 0.005.
A loss coefficient of $1e-7$ is used for the KL penalty.
For \modelnameldm{}, we set $\alpha=5$ and $\beta=1$. 
All training and evaluation is done on an NVIDIA H100 GPU.

\begin{table}[t]
\centering
\caption{Image quality metrics. $\uparrow$: higher is better, $\downarrow$: lower is better}
\label{tab:model-comparison}
\begin{tabular}{l@{\hspace{0.5cm}}r@{\hspace{0.5cm}}r@{\hspace{0.5cm}}r@{\hspace{0.5cm}}r}
\toprule
 & Sex Acc $\uparrow$ & Age MAE $\downarrow$ & FID $\downarrow$ & MS-SSIM $\downarrow$ \\
\midrule
Real & 0.88 & 4.63 & -- & 0.74 \\
\midrule
VAE-GAN & 0.55 & 35.53 & 298.38 & 0.81 \\
$\alpha$-GAN~\cite{kwon2019braingan} & 0.51 & 33.30 & 313.58 & 0.90 \\
VQVAE-AR~\cite{tudosiu2022autoregressive} & 0.62 & 29.62 & 290.01 & 0.80 \\
\midrule
LDM~\cite{pinaya2022brainimaginggenerationlatent} & 0.78 & 12.95 & 273.71 & 0.77 \\
LDM\textsuperscript{c} & 0.79 & 10.37 & 282.52 & 0.79 \\
\modelnameldm{} (ours) & 0.84 & 5.59 & 213.70 & \textbf{0.73} \\
\modelnameldmcond{} (ours) & \textbf{0.85} & \textbf{4.93} & \textbf{202.49} & \textbf{0.73} \\
% MorphGen, displace=0.1 & 0.73 & 6.21 & 277.90 & 0.71 \\
% MorphGen, displace=1 & 0.74 & 7.23 & 240.98 & 0.73 \\
% MorphGen, displace=3 & 0.74 & 6.88 & 237.39 & 0.73 \\
% MorphGen, displace=5 & 0.66 & 7.89 & 239.98 & 0.74 \\
% MorphGen, kl=1e-8 & 0.77 & 8.59 & 204.91 & 0.78 \\
\bottomrule
\end{tabular}
\end{table}
\subsubsection{Baselines.}
% We compare against GAN and LDM-based baselines.
We compare against two GAN-based models, VAE-GAN and $\alpha$-GAN~\cite{kwon2019braingan}, and an autoregressive model (AR) trained on discrete tokens learned through a VQVAE~\cite{tudosiu2022autoregressive}.
We also compare against an LDM model, which minimizes a reconstruction loss $\Lcal_{sim}$ during autoencoder training.
% Note that $\alpha$-GAN was originally proposed for slice-wise generation, but we reimplement it for 3D generation for better comparison.
All architectural and training details are kept as identical as possible between LDM and \modelnameldm{} variants.

All LDM-based models have conditioning information $\bm{c}$ injected in the diffusion UNet $\epsilon_\omega$ via cross-attention with the intermediate feature maps. 
We also experiment with including $\bm{c}$ in the autoencoder by concatenating age and sex as additional channels of $\bm{z}$ repeated across the spatial grid.
For \modelnameldm{}, $\bm{c}$ is also passed as input to the template decoder $\bar{\bm{x}}$.
Throughout this subsection, LDM-based models which have conditioning information injected in the encoder/decoder are denoted with a superscript \textsuperscript{c}.
%we refer to the unconditional variant of \modelname{}LDM (corresponding to Eq.~\eqref{eq:uncond}) as \modelnameldm{} and the conditional variant (corresponding to Eq.~\eqref{eq:cond}) as \modelnameldmcond{}.

\subsubsection{Metrics.}
We report the Frechet inception distance (FID)~\cite{heusel2018ganstrainedtimescaleupdate} using pretrained ImageNet weights, which has been shown to more closely align with expert judgment~\cite{woodland2024fid}.
To quantify diversity, we report the multi-scale structural similarity measure (MS-SSIM)~\cite{pinaya2022brainimaginggenerationlatent,wang2003msssim}, which is computed by averaging over 1000 MS-SSIM values for 1000 pairs of images.
Lower is better since a higher MS-SSIM indicates a higher degree of image similarity and therefore, lower diversity.

To measure how well synthetic samples capture age and sex information, we pass the samples through pretrained models for age and sex and quantify the prediction error.
% Age and sex metadata are included as conditioning information $\bm{c}$ during autoencoder and diffusion training.
We train a CNN-based age regressor and sex classifier on the same (real) data used for training the generative models.\footnote{
The architectures for the age and sex predictors are CNNs with 4 downsampling levels and 2 convolutional blocks (conv, norm, and ReLU) per level.
The final layer is a channel-wise average pooling followed by a fully-connected layer down to a scalar output.
The age regressor is trained to minimize MSE loss and the sex classifier is trained to minimize binary cross entropy.
}

We perform a voxel-based morphometry analysis to evaluate synthetic samples with respect to volumes of brain regions~\cite{Whitwell9661}.
To obtain segmentations for both real and synthetic data, we use SynthSeg~\cite{BILLOT2023102789}, a widely-used tool for segmenting brain MRI volumes.

\begin{figure*}[t!]
  \centering
   \includegraphics[width=1.0\linewidth]{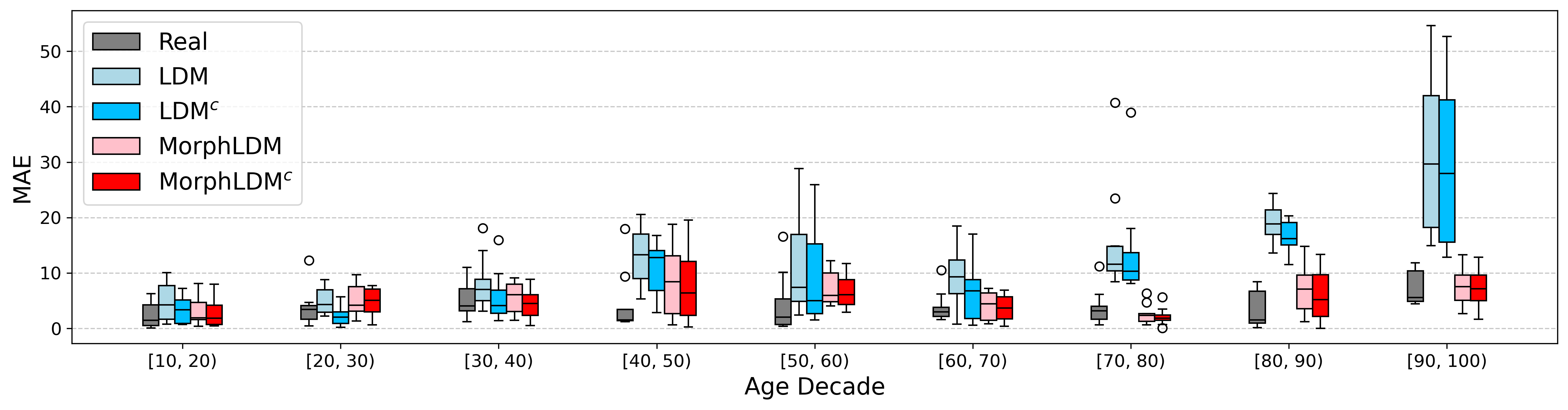}

   \caption{Prediction error of synthetic samples on a pretrained age predictor across age decades. ``Real'' denotes prediction error on real (i.e. validation) data.  
   Generally, we find that \modelnameldm{} models exhibit lower MAE values than the LDM model across all age ranges, especially at older ages, where less training data is available.
   \modelnameldmcond{} generally outperforms \modelnameldm{}.
   }
   \label{fig:age_loss_boxplot}
\end{figure*}
\subsection{Image Generation}
Fig.~\ref{fig:samples} depicts a representative sample across three views for all models.
We observe that LDM-based samples exhibit sharper images, with GAN-based models producing blurry results as well as less diversity across samples.
We observe that \modelnameldm{} has a higher degree of morphological detail compared to LDM, especially in folding patterns in cortical regions.
This may be attributed to the model's ability to focus on morphology instead of other aspects of the image.\footnote{We refer the reader to the Supplementary material for a visualization of the learned templates for \modelnameldmcond{}.}

Table~\ref{tab:model-comparison} summarizes the results on various image generation metrics.
We find that both \modelnameldm{} variants better capture age and sex information, as evidenced by higher sex accuracy and lower age MAE.
% This may be attributed by the fact that age and sex are largely determined by brain structure, and \modelname{} allows for better capturing of structural information in the generation process.
\modelnameldmcond{} generally outperforms \modelnameldm{}, which can be attributed to its increased expressivity in being able to generate age-conditioned templates.
\modelnameldm{} variants also outperform LDM variants on FID and MS-SSIM.
We attribute this to \modelnameldm{}'s ability to capture more detailed and varied morphology compared to LDM.

Fig.~\ref{fig:age_loss_boxplot} plots the age MAE across age decades for all real and synthetic samples.
The improvement of \modelnameldm{} over baseline LDM becomes more pronounced as age increases.
Given that our training dataset is skewed toward brain MRIs of younger subjects (5-20 years), we hypothesize that \modelnameldm{} is more data-efficient due to restricting its modeling to morphology.
% exhibit lower MAE values than the LDM model across all age ranges.
% As before, \modelname{}-cond generally outperforms \modelname{}-uncond, which can be attributed to its increased expressivity in being able to generate a template specific to the age.

% More visualizations of synthetic samples as well as visualizations of learned templates for \modelnameldmcond{} are provided in the Supplementary material.

% In Fig.~\ref{fig:learned_templates}, we visualize the conditional templates learned by the best-performing model variant, \modelname{}-cond.
% Perhaps interestingly, we observe that the learned templates lack any intricate details like folding patterns.
% However, we observe that the definition and size of ventricles 
% We believe that this is a result of the diversity of ages and pathologies present in our training dataset; the model learns that a relatively simplistic template enables the best downstream generaitons.
% Additionally, we believe that adding more constraints to $\Rcal$, thereby restricting the flexibility of the deformations, will likely lead to more realistic templates.
% This is an interesting direction that we leave to future work.

% \begin{figure*}[t]
%   \centering
%    \includegraphics[width=0.7\linewidth]{figures/learned_templates.png}

%    \caption{Learned conditional templates for varying ages.}
%    \label{fig:learned_templates}
% \end{figure*}

\subsection{Effect Size of Regional Volumes}
\begin{figure*}[t!]
  \centering
   \includegraphics[width=1.0\linewidth]{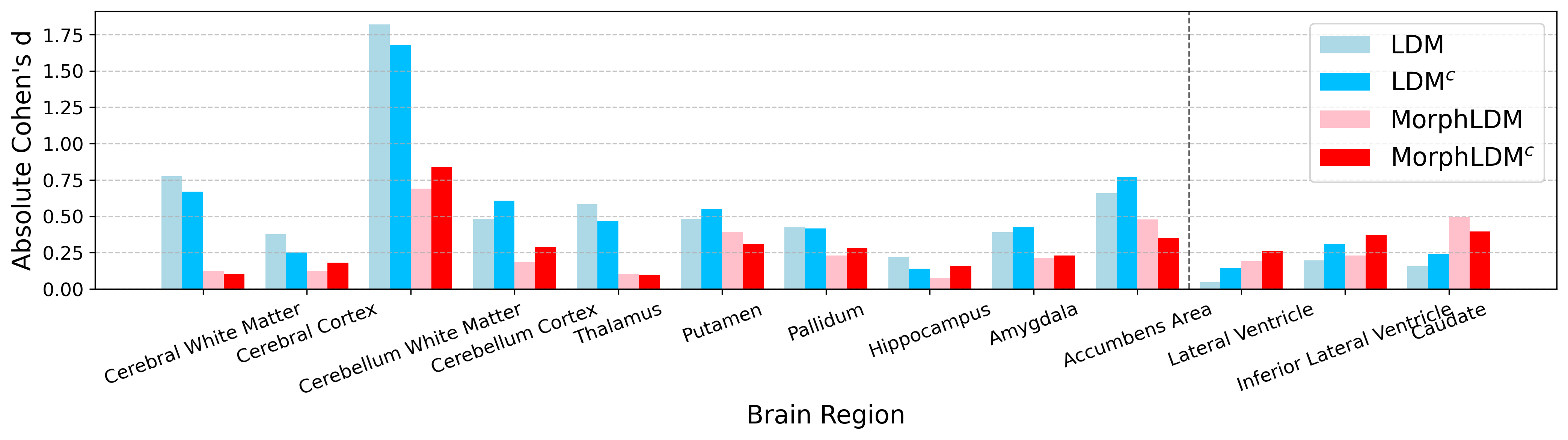}

   \caption{Absolute Cohen's d of regional volumes between real and synthetic samples. Lower is better. \modelnameldm{} variants exhibit lower values for 10 out of 13 regions.}
   \label{fig:cohensd}
\end{figure*}

To quantify the morphological realism of synthetic images, we quantify the effect size of regional volumes between populations of synthetic and real brains~\cite{PENG2024103325,lakens2013effectsizes} using the absolute Cohen's d.\footnote{Absolute Cohen's d is defined as the absolute difference between the two population means divided by the pooled standard deviation:
$
    d = \frac{|\bar{x}_1 - \bar{x}_2|}{s}.
$
}
Fig.~\ref{fig:cohensd} shows a bar graph of absolute Cohen's d values for different models, across all brain regions that are segmented by SynthSeg.
We find that \modelnameldm{} models exhibit lower Cohen's d values for 10 out of the 13 regions (left of the dotted line). 
This indicates a broadly closer degree of anatomical similarity with real samples in terms of regional volumes.

\section{Conclusion}
We propose \modelnameldm{}, a diffusion model for structural brain MRI generation that synthesizes novel displacement fields applied to a learned, universal template.
The ``autoencoder'' of \modelnameldm{} takes as input an image and learnable template and outputs a dense deformation field; this deformation field is applied to the learnable template, and a registration loss is minimized between the original image and the deformed template.
Empirically, our approach outperforms baselines with respect to standard image quality metrics, faithfulness to specified attribute conditions, and voxel-based morphometry.
%
% ---- Bibliography ----
%
% BibTeX users should specify bibliography style 'splncs04'.
% References will then be sorted and formatted in the correct style.
%
% \subsubsection{\ackname} This work was supported in part by National Institutes of Health grant AG089169.
% This study was also supported by the Stanford School of Medicine Department of Psychiatry and Behavioral Sciences Jaswa Innovator Award, the Stanford Institute for Human-Centered AI Postdoctoral Fellowship, and the Stanford HAI Hoffman-Yee Award. 
\subsubsection{\ackname} This work was supported in part by National Institutes of
Health grant AG089169, AA021697, DA057567.  
This study was also supported by the Stanford School of Medicine Department of Psychiatry and Behavioral Sciences Jaswa Innovator Award, the Stanford Institute for Human-Centered AI Postdoctoral Fellowship, the Stanford HAI Google Cloud Credits, and the Stanford HAI Hoffman-Yee Award.

\subsubsection{\discintname}
The authors have no competing interests to declare that are relevant to the content of this article.

%\clearpage
\bibliographystyle{splncs04}
\bibliography{example_paper}

\begin{thebibliography}{10}
\providecommand{\url}[1]{\texttt{#1}}
\providecommand{\urlprefix}{URL }
\providecommand{\doi}[1]{https://doi.org/#1}

\bibitem{BILLOT2023102789}
Billot, B., Greve, D.N., Puonti, O., Thielscher, A., {Van Leemput}, K., Fischl, B., Dalca, A.V., Iglesias, J.E.: Synthseg: Segmentation of brain mri scans of any contrast and resolution without retraining. Medical Image Analysis  \textbf{86},  102789 (2023)

\bibitem{BLAIOTTA2018117}
Blaiotta, C., Freund, P., Cardoso, M.J., Ashburner, J.: Generative diffeomorphic modelling of large mri data sets for probabilistic template construction. NeuroImage  \textbf{166},  117--134 (2018)

\bibitem{chong2021synthesis}
Chong, C.K., Ho, E.T.W.: Synthesis of 3d mri brain images with shape and texture generative adversarial deep neural networks. IEEE Access  \textbf{9},  64747--64760 (2021). \doi{10.1109/ACCESS.2021.3075608}

\bibitem{christensen1997volumetric}
Christensen, G., Joshi, S., Miller, M.: Volumetric transformation of brain anatomy. IEEE Transactions on Medical Imaging  \textbf{16}(6),  864--877 (1997)

\bibitem{dalca2019learningconditionaldeformable}
Dalca, A., Rakic, M., Guttag, J., Sabuncu, M.: Learning conditional deformable templates with convolutional networks. In: NeurIPS. vol.~32 (2019)

\bibitem{dey2022generativeadversarialregistrationimproved}
Dey, N., Ren, M., Dalca, A.V., Gerig, G.: Generative adversarial registration for improved conditional deformable templates (2022)

\bibitem{esser2021tamingtransformershighresolutionimage}
Esser, P., Rombach, R., Ommer, B.: Taming transformers for high-resolution image synthesis (2021)

\bibitem{fernandez2024multipathological}
Fernandez, V., Pinaya, W.H.L., Borges, P., Graham, M.S., Tudosiu, P.D., Vercauteren, T., Cardoso, M.J.: Generating multi-pathological and multi-modal images and labels for brain mri. Medical Image Analysis  \textbf{97},  103278 (2024)

\bibitem{GANSER20043}
Ganser, K.A., Dickhaus, H., Metzner, R., Wirtz, C.R.: A deformable digital brain atlas system according to talairach and tournoux. Medical Image Analysis  \textbf{8}(1),  3--22 (2004)

\bibitem{goodfellow2014generativeadversarialnetworks}
Goodfellow, I.J., Pouget-Abadie, J., Mirza, M., Xu, B., Warde-Farley, D., Ozair, S., Courville, A., Bengio, Y.: Generative adversarial networks (2014)

\bibitem{gulrajani2017improvedtrainingwassersteingans}
Gulrajani, I., Ahmed, F., Arjovsky, M., Dumoulin, V., Courville, A.: Improved training of wasserstein gans (2017)

\bibitem{heusel2018ganstrainedtimescaleupdate}
Heusel, M., Ramsauer, H., Unterthiner, T., Nessler, B., Hochreiter, S.: Gans trained by a two time-scale update rule converge to a local nash equilibrium (2018)

\bibitem{ho2020denoisingdiffusionprobabilisticmodels}
Ho, J., Jain, A., Abbeel, P.: Denoising diffusion probabilistic models (2020)

\bibitem{isola2018imagetoimagetranslationconditionaladversarial}
Isola, P., Zhu, J.Y., Zhou, T., Efros, A.A.: Image-to-image translation with conditional adversarial networks (2018)

\bibitem{jeanneret2022diffusionmodelscounterfactualexplanations}
Jeanneret, G., Simon, L., Jurie, F.: Diffusion models for counterfactual explanations (2022)

\bibitem{Karcher2021}
Karcher, N.R., Barch, D.M.: The abcd study: understanding the development of risk for mental and physical health outcomes. Neuropsychopharmacology  \textbf{46}(1),  131--142 (2021)

\bibitem{kim2022diffusiondeformablemodel4d}
Kim, B., Ye, J.C.: Diffusion deformable model for 4d temporal medical image generation (2022), \url{https://arxiv.org/abs/2206.13295}

\bibitem{kwon2019braingan}
Kwon, G., Han, C., Kim, D.s.: Generation of 3d brain mri using auto-encoding generative adversarial networks. In: MICCAI 2019. pp. 118--126 (2019)

\bibitem{lakens2013effectsizes}
Lakens, D.: Calculating and reporting effect sizes to facilitate cumulative science: a practical primer for t-tests and anovas. Frontiers in Psychology  \textbf{4} (2013)

\bibitem{madan2017advances}
Madan, C.R.: Advances in studying brain morphology: The benefits of open-access data. Frontiers in Human Neuroscience  \textbf{11} (2017)

\bibitem{PENG2024103325}
Peng, W., Bosschieter, T., Ouyang, J., Paul, R., Sullivan, E.V., Pfefferbaum, A., Adeli, E., Zhao, Q., Pohl, K.M.: Metadata-conditioned generative models to synthesize anatomically-plausible 3d brain mris. Medical Image Analysis  \textbf{98},  103325 (2024)

\bibitem{petersen2010alzheimer}
Petersen, R.C., Aisen, P.S., Beckett, L.A., Donohue, M.C., Gamst, A.C., Harvey, D.J., Jack, Clifford~R., J., Jagust, W.J., Shaw, L.M., Toga, A.W., Trojanowski, J.Q., Weiner, M.W.: Alzheimer's disease neuroimaging initiative (adni): clinical characterization. Neurology  \textbf{74}(3),  201--209 (2010)

\bibitem{pinaya2022brainimaginggenerationlatent}
Pinaya, W.H.L., Tudosiu, P.D., Dafflon, J., da~Costa, P.F., Fernandez, V., Nachev, P., Ourselin, S., Cardoso, M.J.: Brain imaging generation with latent diffusion models (2022)

\bibitem{pombo2023equitablemodelling}
Pombo, G., Gray, R., Cardoso, M.J., Ourselin, S., Rees, G., Ashburner, J., Nachev, P.: Equitable modelling of brain imaging by counterfactual augmentation with morphologically constrained 3d deep generative models. Medical Image Analysis  \textbf{84},  102723 (2023)

\bibitem{Pulliam2011}
Pulliam, E., Singleton, A.B.: The parkinson's progression markers initiative (ppmi): Study design and protocol. Movement Disorders  \textbf{26}(9),  1453--1460 (2011)

\bibitem{rombach2022highresolutionimagesynthesislatent}
Rombach, R., Blattmann, A., Lorenz, D., Esser, P., Ommer, B.: High-resolution image synthesis with latent diffusion models (2022)

\bibitem{starck2024diffdefdiffusiongenerateddeformationfields}
Starck, S., Sideri-Lampretsa, V., Kainz, B., Menten, M., Mueller, T., Rueckert, D.: Diff-def: Diffusion-generated deformation fields for conditional atlases (2024), \url{https://arxiv.org/abs/2403.16776}

\bibitem{tudosiu2022autoregressive}
Tudosiu, P.D., Pinaya, W.H.L., Graham, M.S., Borges, P., Fernandez, V., Yang, D., Appleyard, J., Novati, G., Mehra, D., Vella, M., Nachev, P., Ourselin, S., Cardoso, J.: Morphology-preserving autoregressive 3d generative modelling of the brain. In: Simulation and Synthesis in Medical Imaging: 7th International Workshop, SASHIMI 2022, Held in Conjunction with MICCAI 2022, Singapore, September 18, 2022, Proceedings. p. 66–78 (2022)

\bibitem{VANESSEN201362}
{Van Essen}, D.C., Smith, S.M., Barch, D.M., Behrens, T.E., Yacoub, E., Ugurbil, K.: The wu-minn human connectome project: An overview. NeuroImage  \textbf{80},  62--79 (2013), mapping the Connectome

\bibitem{wang2024frameworkinterpretability}
Wang, A.Q., Karaman, B.K., Kim, H., Rosenthal, J., Saluja, R., Young, S.I., Sabuncu, M.R.: A framework for interpretability in machine learning for medical imaging. IEEE Access  \textbf{12},  53277--53292 (2024)

\bibitem{wang2021independent}
Wang, Y., Leiberg, K., Ludwig, T., Little, B., Necus, J.H., Winston, G., Vos, S.B., de~Tisi, J., Duncan, J.S., Taylor, P.N., Mota, B.: Independent components of human brain morphology. NeuroImage  \textbf{226},  117546 (2021)

\bibitem{wang2003msssim}
Wang, Z., Simoncelli, E., Bovik, A.: Multiscale structural similarity for image quality assessment. In: Asilomar Conference. vol.~2, pp. 1398--1402 (2003)

\bibitem{Whitwell9661}
Whitwell, J.L.: Voxel-based morphometry: An automated technique for assessing structural changes in the brain. Journal of Neuroscience  \textbf{29}(31),  9661--9664 (2009)

\bibitem{woodland2024fid}
Woodland, M., Castelo, A., Al~Taie, M., Albuquerque Marques~Silva, J., Eltaher, M., Mohn, F., Shieh, A., Kundu, S., Yung, J.P., Patel, A.B., Brock, K.K.: Feature extraction for generative medical imaging evaluation: New evidence against an evolving trend. In: MICCAI 2024. pp. 87--97 (2024)

\bibitem{yi2019ganinmedicalimaging}
Yi, X., Walia, E., Babyn, P.: Generative adversarial network in medical imaging: A review. Medical Image Analysis  \textbf{58},  101552 (2019)

\bibitem{zheng2024deformationrecoverydiffusionmodeldrdm}
Zheng, J.Q., Mo, Y., Sun, Y., Li, J., Wu, F., Wang, Z., Vincent, T., Papież, B.W.: Deformation-recovery diffusion model (drdm): Instance deformation for image manipulation and synthesis (2024), \url{https://arxiv.org/abs/2407.07295}

\bibitem{zhuo2024diffuseregdenoisingdiffusionmodel}
Zhuo, Y., Shen, Y.: Diffusereg: Denoising diffusion model for obtaining deformation fields in unsupervised deformable image registration (2024), \url{https://arxiv.org/abs/2410.05234}

\end{thebibliography}

\end{document}